# A concise and critical literature review on approaches for the modelling of the deformation behavior of soils subjected to cyclic loads

Anteneh Biru Tsegaye

Senior engineer at the Norwegian Geotechnical Institute

Email: anteneh.biru.tsegaye@ngi.no

**Abstract**: This paper focuses on state-of-the art of various approaches for the modelling of deformation behavior of soils when subjected to cyclic loading. The various approaches are broadly classified into implicit and explicit. Models within each category are discussed and some of their limitations are pointed out.

## 1. Introduction

Various structures that are founded on soils can be subjected to repetitive (cyclic) loads. The loads can vary from a small (in hundreds) of cycles as in the case of earthquake to millions of cycles (as in the case of windfarms). The modelling of deformation behavior of soils subjected to repetitive (cyclic) loading conditions has therefore been a subject of continuous interest in the constitutive modelling community. Such models were needed for instance for design of high way pavements, railway tracks and airports that are subjected to repeated loads and as a result affected by accumulation of excess pore pressures and plastic deformations. Furthermore, alerted by catastrophic events of earthquake-induced failures of buildups on a saturated sand, the modelling of earthquake-induced liquefaction was of a major interest. Usually, earthquakes last for few seconds and can impart large amplitude but mostly relatively few cycles of loads on a foundation soil. The accumulation of pore pressures can then lead to undesired deformations and rotations and further compromising the bearing capacity of the soil which may lead to a complete failure of the overlying structures. Several models have aimed at capturing this cyclic mobility and succeeded to a degree; these models also led to the qualitative understanding of the mechanics of soils when subjected to cyclic loading conditions. Recently, the modelling of deformation behavior of soils subjected to millions of load cycles has become an interest. This is mainly driven by developments in offshore structures that are anchored to the seabed (foundations to platforms at an oil production site) and monopile foundations to windfarms. These structures are subjected to repeated cycles of water waves, wind loads and seismic loads. These developments pose new challenges but also bring new dimensions and new opportunities.

On February 6-7, 2019, the Norwegian Geotechnical Institute (NGI) hosted its first DENSER workshop focusing on the modelling of cyclic mobility of dense sands which prompted this short literature study on the subject.

In the workshop, the various approaches were roughly classified into "implicit (intrinsic)" and "explicit (extrinsic)". The models that were regarded 'implicit (intrinsic)' were constitutive models that aspired to the modelling of the stress-strain response of each cycle while those that were regarded as 'explicit (extrinsic)' were models that focused on the cumulative effects-without directly modelling each cycle.

There is a vast body of literature on the modelling of deformation behavior of soils that are subjected to cyclic loading. In the following, a brief overview of the literature that are deemed sufficient to show the current state-of the art of the topic is presented. The rough classification of the models into implicit (intrinsic) and explicit (extrinsic) that was done at the DENSER workshop suits our purpose. The author's experience is on the implicit kind and hence the focus of this short essay will be mainly on the implicit (intrinsic) modelling approaches. However, for the sake of completeness, some literature from the latter group is also included.

## 2. Implicit approaches

Many of the models/frameworks that aim at the modelling of soil responses when subjected to cyclic loads are based on the theory of plasticity (elastoplasticity). Plasticity theory was originally developed for the modelling of ductile metals. The theory builds on the postulation of decomposition of strains into elastic and plastic and employing various surfaces and evolution rules to accommodate various actions of interest in terms of mechanical response. Earlier elastoplastic models attempted to capture the hysteresis behavior of soils through a paraelastic stiffness formulation, for example Hueckel and Nova[1]. Later developments in the theory of plasticity gave rise to kinematically hardening models that are more conveniently tuned to the modelling of mechanical responses of materials to cyclic loads. One of the kinematic hardening frameworks that is specially adapted to soils and widely used for developing constitutive models for the modelling of deformation behavior of soils to cyclic loads is the bounding surface plasticity. The main architecture of models in the bounding surface plasticity was first laid down in the early 1980's (Dafalias and Herrmann [2], Dafalias [3, 4]). In this framework, the evolution of various state and internal variables and hence accumulation of plastic strains is dictated by the distance of the current state in relation to an image state on some well-defined surfaces. Recent models for soils in the framework of bounding surface plasticity contain at least three such surfaces: a bounding surface that encloses all accessible stress states, a dilatancy surface that defines the phase transformation boundary from a contractive to a dilative state and a yield surface which contains the current stress state and encloses an elastic neighborhood [5]. The evolution of the state variables is then defined based on their distance from their image on from the bounding and the dilatancy surfaces. The bounding surface plasticity framework is a more geometric and an easy to visualize framework. The well-known concepts of critical state and stress-dilatancy are easily accommodated into the framework. Many soil models developed in this framework captured the essential features of observed stress-strain behaviors of soils in triaxial and simple shear conditions. They also captured essential features of responses of soils to cyclic loading. However, many models developed in this framework suffered from excessive ratcheting behavior, i.e., relatively high accumulation of volumetric strains (pore pressures under undrained conditions) in only few cycles. The shear stress-shear strain plot shows an unrealistic saw-tooth trace (called also ratcheting) under the application of repeated unloading and reloading schemes. As a remedy to this unrealistic behavior, researchers have now proposed the introduction of a memory surface that accounts for fabric effects [5]. The memory surface controls the stiffness of the model in loading ranges the soil has previously experienced. The addition of such a memory surface led to a significant improvement over the previous models. Bounding surface plasticity models are also beginning to include effects of principal stress rotations [6].

An alternative arrangement, which can also be used for the modelling of deformation behavior of soils subjected to cyclic loads, involves the use of multiple yield loci that are nested together [7]. Each locus has a simple hardening law associated with it and the hardening law determines the contribution of the specific yield loci to the overall plastic strain [8]. The advantage of such models is that they do not require complex functions to be chosen. Their downside however is that they required many material constants.

Another approach for modelling the hysteresis behavior of soils subjected to cyclic loads is Iwan's approach [9]. Iwan proposed what he called a distributed element model, coupling of simple models in series and in parallel. Iwan's approach has been used by some researchers, for instance [10], and its simplicity is appealing. How realistic the hysteresis behavior and to what extent the response depends on the formulation of the constitutive model needs to be investigated.

Cyclic plasticity models can also be formulated in a multiplane (and multiple spring) frameworks [11-13]. These frameworks simplify the complexity of the various constitutive equations as they are designed to enable the integration of a 3-dimensional response from 1D/ 2D formulations. They are also naturally suited for including directional variation of material properties and modelling of principal stress-rotations. In these frameworks, stress increments and strain increments are transformed into planes; the non-linear responses obtained at the planes are back transformed into their respective global spaces. Nishimura and Towhata [14] employed the multispring framework for developing a cyclic elastoplastic model for frictional materials. Nishimura [15] used an icosahedron geometry for distributing planes that host "springs". Six springs are distributed on each plane. Constitutive relations between stresses and strains are formulated for each spring. To capture the hysteresis behavior of soils, an extended Masing's rule is encoded into the formulation of the shear stress-shear strain relationship of the constituent one-dimensional shear mechanism. Furthermore, a linear relationship is written down between the stress ratio and dilatancy ratio. Element simulations of deformations under cyclic loading in a simple shear condition have shown that the model is capable of reproducing some of the qualitative aspects of cyclic mobility under such a set-up. A closer investigation is required to check whether the model is suitable for the modelling soil responses when subjected to high number of load cycles and varying load magnitude. These frameworks are promising. However, they can be computationally costly.

The models and frameworks presented so far give some idea of the various ways in which the plasticity theory is employed for the modelling the response of soils subjected to cyclic loading. Although the various approaches can involve rather complex mathematics, most of the ingredients are arbitrarily selected. For avoiding possible non-physical response, the laws of thermodynamics should be invoked as additional constraints. Since the 1980s, the thermomechanical plasticity framework surfaced as an alternative and a more formalized approach to plasticity theory of soils. Models in the thermomechanical framework begin by assuming an isothermal condition and establishing a free energy function and a dissipation function [8, 16]. The entire constitutive model is then derived from the two functions. The Legendre transform has become quite instrumental in their derivation. This gives an easier route to obeying the laws of thermodynamics contra selecting arbitrary elastoplastic ingredients (yield functions, hardening rules, and dilatancy rules) and later invoking the laws of thermodynamics. In due course, thermomechanical approaches accommodated non-associated plasticity. Soils exhibit complicated memory storing and memory erasing mechanics when subjected to cyclic loading and finding thermodynamic potentials that are suitable for capturing the such a complex behavior can be a challenging task.

Other competing models that are continuously probing into the arena of soil modeling are models within the hypoplasticity framework. The hypoplasticity framework began in the late 70's as an alternative approach to the modeling of mechanical behavior of soils, independently in Germany and in France [17-19]. Hypolasticity models have some features that makes them distinct from elastoplastic models. They are formulated by selecting functions from the general representation theorem of tensor valued isotropic functions of stresses and strain rates. To get the desired structure, various conditions are imposed. The earlier versions of hypoplasticity did not make reference to any kind of surface and no loading and unloading criteria were imposed. Neither is the decomposition of the strain into elastic and plastic necessary. There exist several versions of hypoplasticity for both sands and clays [20-22]. Their application to cyclic loading shows that they capture some essential features but generally perform poorly [20, 21]. They suffered from excessive ratcheting. To tackle this problem, additional state variable was introduced. The new state variable was then called Intergranular Strain (IS). The IS was applied as an overlay onto hypoplastic models to reduce their excessive ratcheting behavior when they are subjected to cyclic loads. As a result, the hypoplastic models with the IS overlay could show some improvement regarding the unrealistic accumulation of strains (and pore pressures). The IS addition came with its own limitations. The formulation is non-rigorous and disobeys a theorem in isothermal cyclic process, which says, "*In every isothermal process starting from a state of equilibrium, the total stress work around a closed path is non-negative.*" This could, in some instances, lead to overshooting of strength during small strain cycles. It may also lead to a non-physical response (may create energy) when a large step unloading is performed following several small strain cycles. The original formulation of the IS was isotropic. Recently, an anisotropic version of the Intergranular Strain (AIS) was developed [23]. The AIS overlay showed an improved response for the simulation of anisotropically constituted soil samples subjected to cyclic loads. However, the extension does not seem to address the other issues mentioned of the original IS formulation. Recent developments due to Wu *et al*. [24] are tending to indicate that the widely used hypoplasticity model for sands that was developed by von Wolffersdorff [21] might have had "flaws in its formulation" and may not have generally been able to preserve the critical state at large strains. To what extent this might have added to the limitation of the model in the modeling cyclic mobility is not yet investigated.

A recent addition to the family of hypoplastic models is Barodesy [25]. The Barodesy framework was originally developed for sands and recently for clays. According to Kolymbas, Barodesy is an evolution equation where the stress rate is expressed as a tensorial function of stress, stretch and other parameters like void ratio. However, closely looking at loading-unloading drained triaxial simulations using the model, it is observed that the response of Barodesy deviates from the experimentally observed. This can be due to some limitation in the implicit formulation of the stress-dilatancy relation. The model may therefore need to be further appropriated for realistic modelling of cyclic mobility of dense sands.

## 3. Explicit approaches

Yet, there are approaches that attempt the modeling of deformation responses of soils under cyclic loading without directly modelling each cycle. Such models were the ones referred to as explicit models in the DENSER workshop. Explicit stiffness degradation methods and pore pressure accumulation methods are mainly built on directly expressing the respective variables as a function of cycle number. The accumulated strains are then related to the cycle number. Explicit methods follow a much more simplified approach to the modelling of deformation behavior of soils subjected to cyclic loading. There are various explicit models for stiffness degradation, for pore pressure generation, and for more advanced plastic strain accumulation. Due to their lumpsum nature, better control may be achieved on the accumulation of plastic strains under high number of cycles.

Niemunis *et al.* [26] developed a high cycle ($N>10^4$) explicit model for accumulation of strains in sand due to cyclic loads of small magnitude (where strain amplitudes of less or equal to one in thousand are considered small and number of cycles greater than 10000 are considered high number of cycles). The model uses the hypoplastic model by Wolffersdorff [21] for low cycles and an explicit accumulation model for high cycles. More and more of this type of models are being developed recently. For example, the UDCAM (Undrained Cyclic Accumulation Model) and PDCAM (Partially Drained cyclic Accumulation Model) by Jostad *et al.* [27, 28] fall in this category. UDCAM uses 3D strain contour diagrams from undrained cyclic and monotonic triaxial and DSS tests. The model accounts for cyclic degradation by using the cyclic strain accumulation procedures developed at the NGI in the seventies and idealizing the cyclic load as a composition of load portion with constant average and cyclic loads in each portion. The model was verified by back calculating a model test of a gravity base structure (GBS) in soft clay subjected to monotonic and cyclic loading. The PDCAM is an extended version for considering effect pore pressure dissipation.

A foreseeable benefit of the models of the explicit type is that they require less computation time and numerical troubles can be relatively minimized. However, the author believes they offer less insight into the behavior of the soil; they are difficult to generalize and cycle number as an internal variable may not be fully objective.

## 4. Summary

The deformation behavior of soils under cyclic loading is complex and recent developments in both offshore and onshore increased the need for a capable model. There are several approaches that are tuned towards the modeling of cyclic mobility and cyclic liquefaction behavior of soils with varying degree of success. The approaches were broadly classified into implicit and explicit. The former attempt to model each cycle and the latter are based on cycle number as an internal variable. The implicit approaches have some appeal in them as they engage in the modelling of the detailed processes in the deformation behavior of soils subjected to cyclic loading. Whereas the explicit approaches consider stiffness degradation and pore pressure accumulation based on the cycle number while at the same time decomposing the loading into an average and a cyclic component. Both approaches have their merits and demerits and the emphasis of this essay is on the former. Accordingly, various ways in which the theory of elastoplsaticty can be employed are discussed; some benefits of thermodynamic considerations are pointed out; and alternative modelling approaches in the hypoplasticity framework are briefly discussed and their limitations are pointed out.

**Acknowledgement**: The author would like to acknowledge Multiconsult Norge AS for making the resources available for the author to attend the first DENSER workshop.